\documentclass[epj]{svjour}
\usepackage{epsfig, amsmath}
\begin{document}
\title{What does the $\rho$-meson do? In-medium mass shift scenarios versus hadronic model calculations}
\author{J\"org Ruppert\inst{1} 
, Thorsten Renk\inst{2} \inst{3}
}                     
\offprints{}          
\institute{ Department of Physics, Duke University, PO Box 90305, Durham, 27708, NC, USA \and Department of Physics, PO Box 35 FIN-40014 University of Jyv\"askyl\"a, Finland \and Helsinki Institut of Physics, PO Box 64 FIN-00014, University of Helsinki, Finland}
\date{Received: date / Revised version: date}
%
\abstract{The NA60 experiment has studied low-mass muon pair production in In-In collisions at $158 {\rm AGeV}$ with unprecedented precision. With these results there is hope that the in-medium modifications of the vector meson spectral function can be constrained more thoroughly than before. We investigate in particular what can be learned about collisional broadening by a hot and dense medium and what constrains  the experimental results put on in-medium mass shift scenarios. 
The data show a clear indication of considerable in-medium broadening effects but disfavor mass shift scenarios where the $\rho$-meson mass scales with the square root of the chiral condensate. Scaling scenarios which predict at finite density a dropping of the  $\rho$-meson mass that is stronger than that of the quark condensate are 
clearly ruled out since they are also accompanied by a sharpening of the spectral function. 
} 

\authorrunning{J. Ruppert, T. Renk}
\maketitle
\section{Introduction}
\label{intro}

The properties of vector mesons, especially the $\rho$-meson are expected to change 
in the a hot and dense nuclear medium as created in ultra-relativistic heavy ion collisions.
The CERES and HELIOS collaborations studied dilepton production in central nucleus-nucleus
collisions and observed a strong enhancement of the emission of low-mass dileptons
compared to scaled proton-nucleus and proton-proton collisions \cite{Masera:1995ck,Agakishiev:1995xb,Agakishiev:1998vt,Lenkeit:1999xu}.

It has been argued that these experimental findings could be an indication of chiral symmetry restoration. The approximate chiral symmetry of Quantum Chromodynamics (QCD) is spontaneously broken in the vacuum but expected to be restored at high temperatures or densities.  Indeed lattice QCD calculation predict that the properties of strongly interacting matter change drastically at a temperature about $T_c \approx 170 {\rm MeV}$ above which chiral symmetry is restored and quarks and gluons are 
deconfined. These calculations also indicate that this transition is (at zero net baryon density) a rapid crossover from the low-temperature hadron dominated phase into the high-temperature plasma phase.  For a review, see e.g. \cite{Laermann:2003cv}.

A gradual restoration of chiral symmetry should also lead to modifications in the hadronic
sector (at high densities and temperatures), since chiral partners become degenerate once chiral symmetry is fully restored. Chiral symmetry especially dictates a relationship between the vector and axial-vector channels \cite{Weinberg:1967kj,Kapusta:1994gi}. 
This is important for the dilepton production in a 
 medium since the vector correlator dominates the electromagnetic spectral function. Therefore it is expected that chiral symmetry restoration should manifest itself in modifications of the properties of the $\rho$, $\omega$ and $\phi$ which 
leave its trace in the measured dilepton spectra.

Brown and Rho proposed that chiral symmetry restoration should primarily be reflected in a scaling of the masses of the vector mesons in a nuclear medium with the quark condensate \cite{Brown:1991kk,Brown:1995qt,Brown:2001nh,Brown:2002is}.
This would cause the maximum of the $\rho$-meson spectral function to be shifted
to lower masses in a hot and dense nuclear medium and thus would lead to an enhancement of the lepton pairs produced at lower invariant masses as compared to the vacuum.

We also mention that Brown-Rho scaling is not the only conjecture that indicates substantial mass shifts. A different approach is based on the assumption that the lower-mass meson multiplets should be the relevant degrees of freedom near the phase boundary. Pisarski proposed to study a gauged linear $\sigma$-model to gain insight into vector meson modifications at finite temperature \cite{Pisarski:1994yp,Pisarski:1995xu,Pisarski:1995ck}. He found in a finite temperature mean-field calculation assuming vector meson dominance that the $\rho$-meson and $a_1$-meson masses become degenerate at $T_c$ resulting in $m_{\rho}=m_{a_1}=962 ~{\rm MeV}$. This result is sensitive to the vector-meson dominance conjecture,
since a situation where vector meson dominance does not hold leads to slightly  
decreasing masses close to  the phase transition  $T_c$, namely $m_{\rho}=m_{a_1}=629~{\rm MeV}$.

An alternative picture to these scenarios, based on broadening of the $\rho$-meson without a substantial mass shift of the $\rho$-meson due to interactions with the surrounding thermal medium was suggested by  the study of effective Lagrangians in hadronic model calculations  (see e.g. \cite{Rapp:1999ej,Alam:1999sc,Gale:2003iz}). These findings are also supported by calculations of the spectral function as inferred from experiment by constructing scattering amplitudes 
for vector mesons scattering from pions and nucleons \cite{Eletsky:2001bb,Martell:2004gt}.
This substantial broadening without strong mass shifts of the vector mesons can also be in accord with the gradual  restoration of chiral symmetry if the axial-vector channel exhibits congruently features.

The mass resolution of the dilepton spectra of the CERES experiment were not high enough to conclusively decide which of the scenarios is realized in nature \cite{Rapp:1999ej}. The high precision di-muon data for $158 {\rm AGeV}$ In-In collisions as measured by the NA60 collaboration \cite{NA60data} now provide a possibility to put stronger constraints on our understanding of in-medium effects on vector-mesons. 
Towards this aim we conduct a comparison of different model calculations with data. 

Our strategy is as follows: 
Ideally, one would need
to construct an evolution scenario for In-In collisions that is directly related to hadronic observables in In-In collisions,
however in practice these are not yet sufficiently well determined experimentally. 
This is why we rely on scaling arguments to make a sophisticated attempt at describing the evolution of matter
in $158 {\rm AGeV}$ In-In collisions based on an evolution model for Pb-Pb collisions which has been 
successfully used to describe the plethora of observables in Pb-Pb collisions \cite{SPS}.

 Using this evolution model, we then calculate the resulting dilepton spectra for different scenarios of in-medium 
 modifications of vector mesons and compare with data.  
 We are specifically interested in what the NA60 data can tell about dropping mass scenarios and broadening effects. 
 
\section{The evolution model}
A detailed description of the evolution model is found in \cite{SPS}. We restrict ourselves to a brief discussion here.  The scaling from {158 {\rm AGeV}} Pb-Pb to In-In collisions is discussed in more detail in \cite{Renk:2006ax,Renk:2006dt}. 

The evolution model's main assumption is that 
an equilibrated system is created at a short time $\tau_0$ after the onset of the collision. Furthermore, 
it is assumed that the thermal fireball expands isentropically until the mean free path of the 
particles exceeds the dimensions of the systems (at a freeze-out time 
$\tau_f$).

We use the following ansatz for the entropy density:
\begin{equation}
s(\tau, \eta_s, r) = N R(r,\tau) \cdot H(\eta_s, \tau)
\end{equation}
with $\tau $ the proper time as measured in a frame co-moving
with a given volume element  and $R(r, \tau), H(\eta_s, \tau)$ two functions describing the shape of the distribution
and $N$ a normalization factor.They are chosen to be Woods-Saxon distributions
\begin{equation}
\begin{split}
&R(r, \tau) = 1/\left(1 + \exp\left[\frac{r - R_c(\tau)}{d_{\text{ws}}}\right]\right)
\\ & 
H(\eta_s, \tau) = 1/\left(1 + \exp\left[\frac{\eta_s - H_c(\tau)}{\eta_{\text{ws}}}\right]\right).
\end{split}
\end{equation}
to describe the shapes for a given $\tau$. 

The ingredients of the model are therefore the 
skin thickness parameters $d_{\text{ws}}$ and $\eta_{\text{ws}}$
and the para\-me\-tri\-zations of the expansion of the spatial extensions $R_c(\tau), H_c(\tau)$ 
as a function of proper time. For a detailed account how those are parameterized 
see  \cite{Renk:2006ax,Renk:2006dt}.

The parameters of the model have been adjusted to hadronic transverse mass spectra, HBT correlation measurements in $158 {\rm AGeV}$ Pb-Pb and Pb-Au collisions at SPS. Since at this point a freeze-out analysis or HBT
correlation data are not available from NA60 we rely on scaling arguments.

The change in the total entropy production can be inferred  from the ratio of charged particle rapidity densities
$dN_{ch}/d\eta$ at 30\% peripheral Pb-Au collisions with $2.1 < \eta <2.55$ measured by CERES \cite{Agakishiev:1995xb,Agakishiev:1998vt,Lenkeit:1999xu} and semi-central In-In collisions at $\eta=3.8$ measured by NA60 \cite{NA60data}, $dN_{ch}^{In-In}/dN_{ch}^{Pb-Pb} = 0.68$.

The number of participant baryons and the effective initial radius are obtained with nuclear overlap calculations. We assume that the stopping power scales approximately with the number of binary collision per participant, $N_{bin}/N_{part}$. The shape parameters $d_{ws}$ and $\eta_{ws}$ have no great impact on electromagnetic emission into the midrapidity slice, they primarily govern the ratio of surface to volume emission of hadrons. Thus, we leave them unchanged from their value determined in \cite{SPS}.
Under the assumption that the
physics leading to equilibration is primarily a function of incident energy we keep the formation time $\tau = 1$ fm/c as in
Pb-Pb collisions. We stress that the final results exhibit no great sensitivity to the choice of either
$\eta_0$ or $\tau_0$ except in the high $M$, high $p_T$ limit.

The largest uncertainty is the choice of the decoupling temperature $T_F$. Due to the smaller system size of 
In-In, a higher decoupling temperature can be expected. An unambiguous answer could be
obtained from HBT correlations and the transverse mass spectra, for the time being we tentatively choose $T_F=130 {\rm MeV}$. This translates primarily into an uncertainty in the normalization of the results.
The flow velocity is fixed at $T_C$ by choosing the value of transverse flow obtained at $T=130 {\rm MeV}$ in the 
Pb-Pb model in the same centrality class. 

There is likewise an uncertainty (which is to some degree correlated with the decoupling temperature) since flow determines the slope of hadronic (and to a lesser degree electromagnetic) $p_T$ spectra.

The equation of state in the hadronic phase and the off-equilibrium parameters $\mu_\pi, \mu_K$ are inferred from statistical model calculations as described in \cite{SPS,Hadronization}. The resulting fireball evolution is characterized by a peak temperature of about $250$ MeV, a lifetime of $\sim 7.5$ fm/c and a top transverse flow velocity of $0.5 c$ at decoupling.

\section{The hadronic spectral function}
Different effective hadronic models and techniques for the calculations of the properties of hadronic
matter near and below the phase boundaries have been suggested. for reviews of the intense theoretical activities,
see e.g. \cite{Rapp:1999ej,Alam:1999sc,Gale:2003iz}. 
Most of those calculations predict substantial broadening in matter with comparably small shifts of the 
in-medium masses. However, this might be different if one follows the conjecture by Brown and Rho that in-medium vector-meson masses decrease considerably due to the gradual restoration of chiral symmetry and a scaling of the masses with the chiral condensate in the vicinity of the phase transition \cite{Brown:1991kk,Brown:1995qt,Brown:2001nh,Brown:2002is}. We therefore discuss this issue
separately from hadronic model calculations. 

\subsection{Hadronic model calculations}
We briefly discuss three different ha\-dronic model calculations which predict
broadening effects but small mass shifts as generic examples. 
The first model is based on \cite{RSW,SW} where the electromagnetic current-current correlator  has been computed in the so called improved vector meson dominance model 
combined with the chiral dynamics of pions and kaons.
We refer to this model as  'chiral model' further on. 
For the evaluation of finite baryon density effects which are relevant at SPS conditions the results from Klingl, Kaiser, and Weise \cite{Klingl:1997kf} are used. Thermal broadening of the $\rho$, $\phi$, and $\omega$ were calculated by Schneider and Weise as modifications on top of the finite baryon density effects using perturbative methods \cite{SW,RSW}.
 In these calculations it is assumed that the temperature and density dependences of the vector meson self-energy factorize. This amounts to neglecting contributions from pion-nucleon scattering where the pion comes from the heat baths. Formally speaking matrix element contributions such as $\left< \pi N | {\cal T} j_\mu(x) j^{\mu}(x) |\pi N \right>$ are neglected.

A second approach \cite{Renk:2003hu} 
(referred to as 'Walecka model' in the following) is based on a different idea: The method of thermofield dynamics is used to calculate the state with minimum thermodynamic potential at finite temperature and density. The temperature and density dependent baryon and sigma masses are calculated  selfconsistently.
The medium modification to the masses  of the $\omega$- and $\rho$-mesons in hot nuclear matter including the quantum correction effects are then calculated in the relativistic random phase approximation. The decay widths for the mesons are calculated from the imaginary part of the self energy using the Cutkosky rule.

We also discuss a third model calculation where we solved truncated Schwinger-Dyson
equations incorporating a self-consistent resummation of the $\pi-\rho$ interaction in order to determine the thermal broadening of the $\rho$-meson, see \cite{RR} and especially note \cite{Note}. We refer to this model as the $\pi-\rho$ model. 
This $\Phi$-functional approach self-consistently takes into account the finite in-medium damping width of the pion in the thermal heat bath. Effects of baryons are not yet included in the self-consistent approximation scheme. The solution of the $\Phi$-functional approach includes the full momentum dependence of the three-dimensional longitudinal and transverse components of the $\rho$-meson spectral  
function which is especially important for a comparison with the NA60 data measured at integrated and high $p_T$, see \cite{Renk:2006dt} section II D. 

\subsection{Mass scaling model}
\label{mass-scale}
The universal scaling of the masses of hadronic particles, especially for the $\rho$-meson with the chiral
condensate were proposed by Brown and Rho \cite{Brown:1991kk}. 
Given this conjecture, the gradual restoration of chiral symmetry would cause the maximum of the spectral function of the $\rho$-meson to be shifted to lower masses in heavy-ion collisions. This would lead to an enhancement of lepton 
pair production at lower invariant masses as compared to the vacuum.
Calculations in the hidden-local-symmetry model at finite temperature indeed support the notion of dropping masses \cite{Harada:2001it,Harada:2005br}.
But an explicit calculation of spectral functions in such an approach, including finite temperature and baryon 
density as well as momentum dependence of the spectral function is not yet available. 

Therefore we take a different approach here and restrict ourselves to the discussion of simplified schematic scaling models.

We construct a simple mass scaling model which allows for the investigation of a dropping of 
the $\rho$-mass scaled with  the quark condensate and takes in-medium changes of the width 
of the $\rho$-meson in a nuclear medium into account. 

This model has three main assumptions:
i) the relative change of the in-medium mass of the $\rho$-meson is proportional to the relative drop of the quark condensate to some power $a$ in the medium:
\begin{eqnarray}
\label{scaling}
\frac{m^*_{\rho}}{m_{\rho}}=\left(\frac{\langle q \bar{q} \rangle^*}{\langle q \bar{q} \rangle}\right)^a,
\end{eqnarray}
(ii) the width of the $\rho$-meson can change in the medium, and (iii) the validity of vector meson dominance is assumed.
For simplicity we assume the rho-meson to be at rest and employ a Breit-Wigner parametrization for the $\rho$-meson spectral function \cite{Leupold:1997dg,BR} and note that this ansatz may not account for peculiar features, e.g. two separate pole structures, that could appear in certain classes of hadronic models.

Taking into account the change of the width of the $\rho$-meson is necessary since predictions for the in-medium mass and in-medium width of the $\rho$-meson in a nuclear medium cannot be independent but have to be strongly correlated in order to fulfill the QCD sum rule approach \cite{BR,Leupold:1997dg}.

For simplicity we restricted our analysis to QCD sum rules at finite
baryon density only \cite{BR} and did not include temperature effects on the condensate/spectral function. 
Although within the $(T,\mu_B)$ trajectory probed by the evolution in the In-In system baryon density 
seems to be the leading influence on the condensate, our model should only be taken as a lower limit of the potential downward mass scaling effect, the actual mass shift realized in certain space-time regions of the collision may be stronger. 

In \cite{BR} it was shown that the lower and upper bounds of the in-medium width of the $\rho$-meson
at rest that are compatible with the QCD sum rules depend on how rapidly the mass decreases in 
comparison to the change of the quark condensate. The bounds for the in-medium
width increase with density if the relative change of the quark condensate is stronger than the relative decrease in mass \cite{BR}.

Assuming a specific scaling law one can extract the upper and lower bounds of the in-medium width of the 
$\rho$-meson at a given density. Those bounds are sensitive to the parameter sets for the QCD
sum analysis \cite{BR}.  Differences between the bounds for different parameter sets show intrinsic uncertainties of the approach. Using these boundaries for the in-medium width one
can study the question if the NA60 data are compatible with a specific scaling law of the in-medium
$\rho$-meson mass. 

To be specific, we consider here the case a=$1/2$ in \cite{BR} (see Fig. 1 therein). We analyzed the upper and lower bounds of compatible width in this case for three different parameter sets of the QCD sum rule approach \cite{Hatsuda:1991ez,Klingl:1997kf,Leinweber:1995fn}.   The case $a=1/2$  has some resemblance to the
Brown-Rho scaling law \cite{Brown:2001nh,Brown:2002is} if one identifies the parametric mass of Brown-Rho scaling with the in-medium $\rho$-meson mass \cite{Brown:2005ka,Brown:2005kb}. This could be a reasonable approximation up to about normal nuclear matter density, $\rho_B=\rho_0$, since the dense loop term contribution is expected to be small \cite{Brown:2005ka,Brown:2005kb}. 
In spite of these obvious similarities between our mass scaling scenario and the Brown-Rho scaling law one should be cautious and consider the arguments by Brown and Rho \cite{Brown:2005ka,Brown:2005kb} why an identification
of Brown-Rho scaling with simplified mass scaling models is problematic. 
They propose to address the scaling issue (especially also the identification of parametric and in-medium $\rho$-meson mass) in the framework of the hidden local symmetry with vector manifestation \cite{Brown:2005kb,Harada:2001it,Harada:2005br}.
Although the issue of a possible violation of vector meson dominance has been stressed recently, this question cannot be expected to be of primarily importance for an understanding of the lineshape of the
spectra but would effect the overall normalization. 

\section{Results}

\begin{figure*}[!htb]
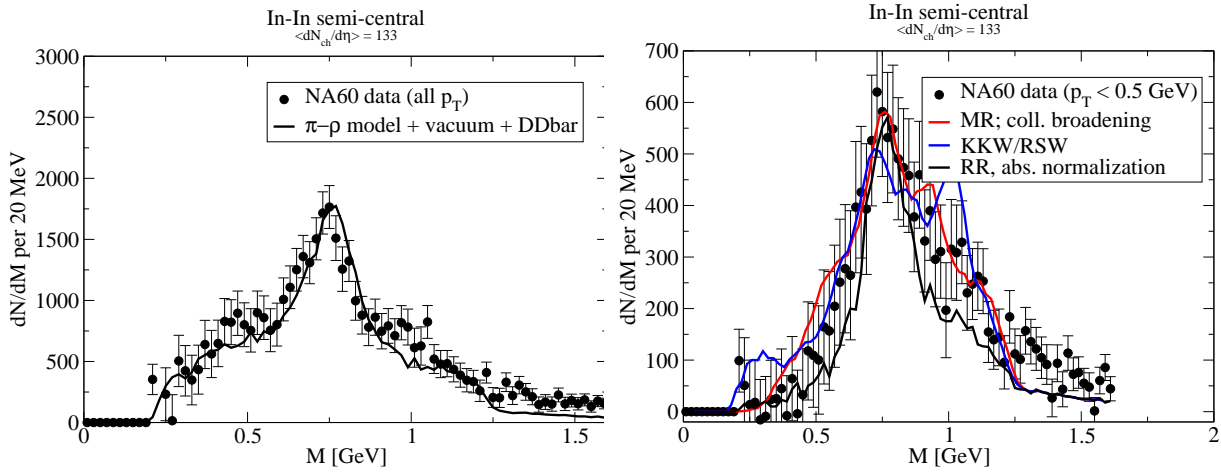


\epsfig{file=dileptons-InIn-RR2-allp_T.eps, width=8cm}
\epsfig{file=dileptons-InIn-MR-RSW-vac-lowp_T2.eps, width=8cm}
\caption{\label{F-RR} 
Left panel: Comparison of the NA60 dimuon data 
\cite{NA60data} with calculations within a self-consistent $\Phi$-functional approach for the $\pi-\rho$ interaction \cite{RR} with the all $p_T$-data.
Right panel:Comparison of the NA60 dimuon data \cite{NA60data} within a $\Phi$-functional approach for the $\pi-\rho$ interaction (RR) \cite{RR}, a chiral model ('RSW/KKW') \cite{RSW} and a calculation in the Walecka model (MR) \cite{Renk:2003hu}.  Model calculations for the chiral model and Walecka model have been normalized to the data in the mass interval $M < 0.9$ GeV.}
\end{figure*}

\begin{figure*}[!htb]
\epsfig{file=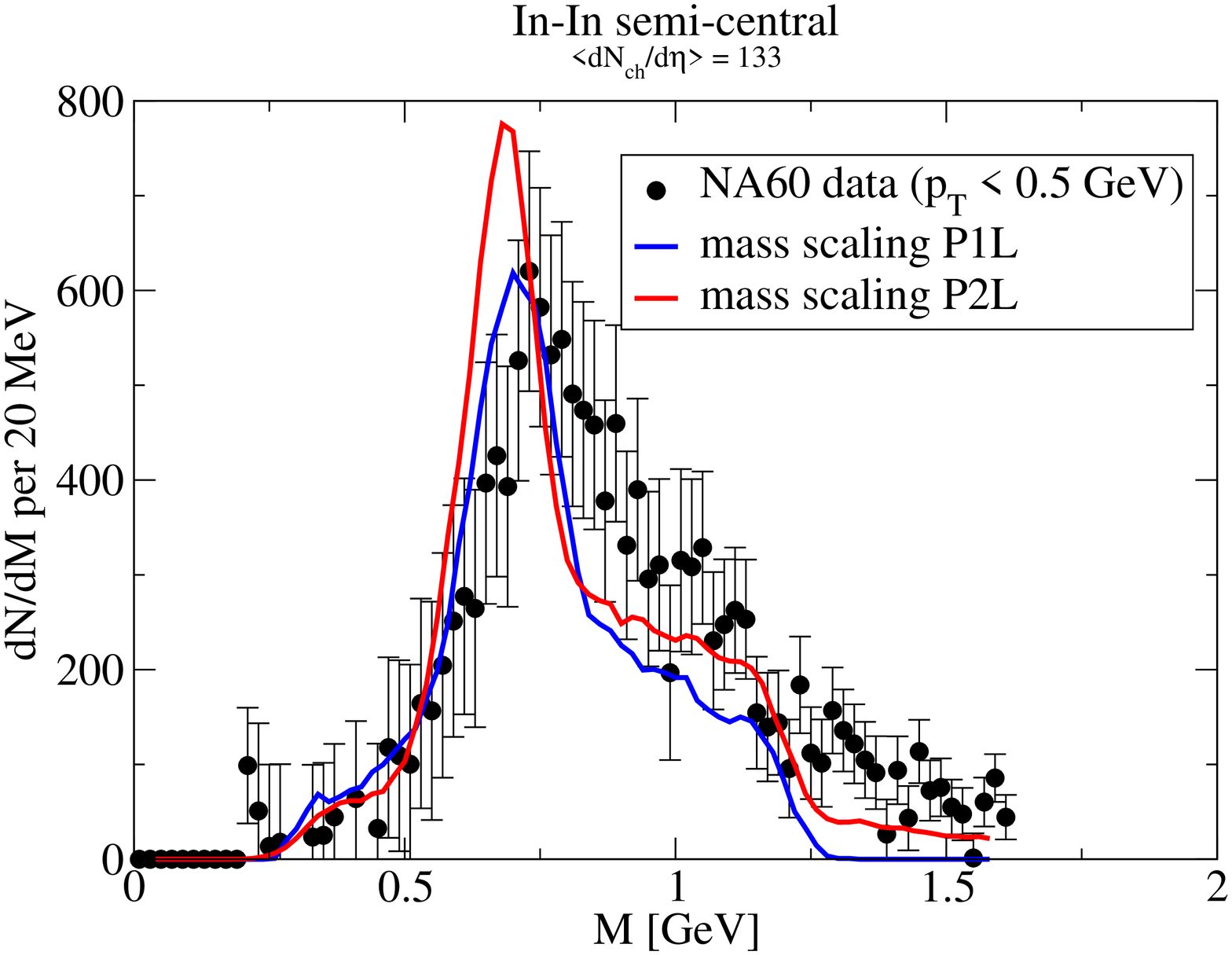, width=7cm,angle=-90}\epsfig{file=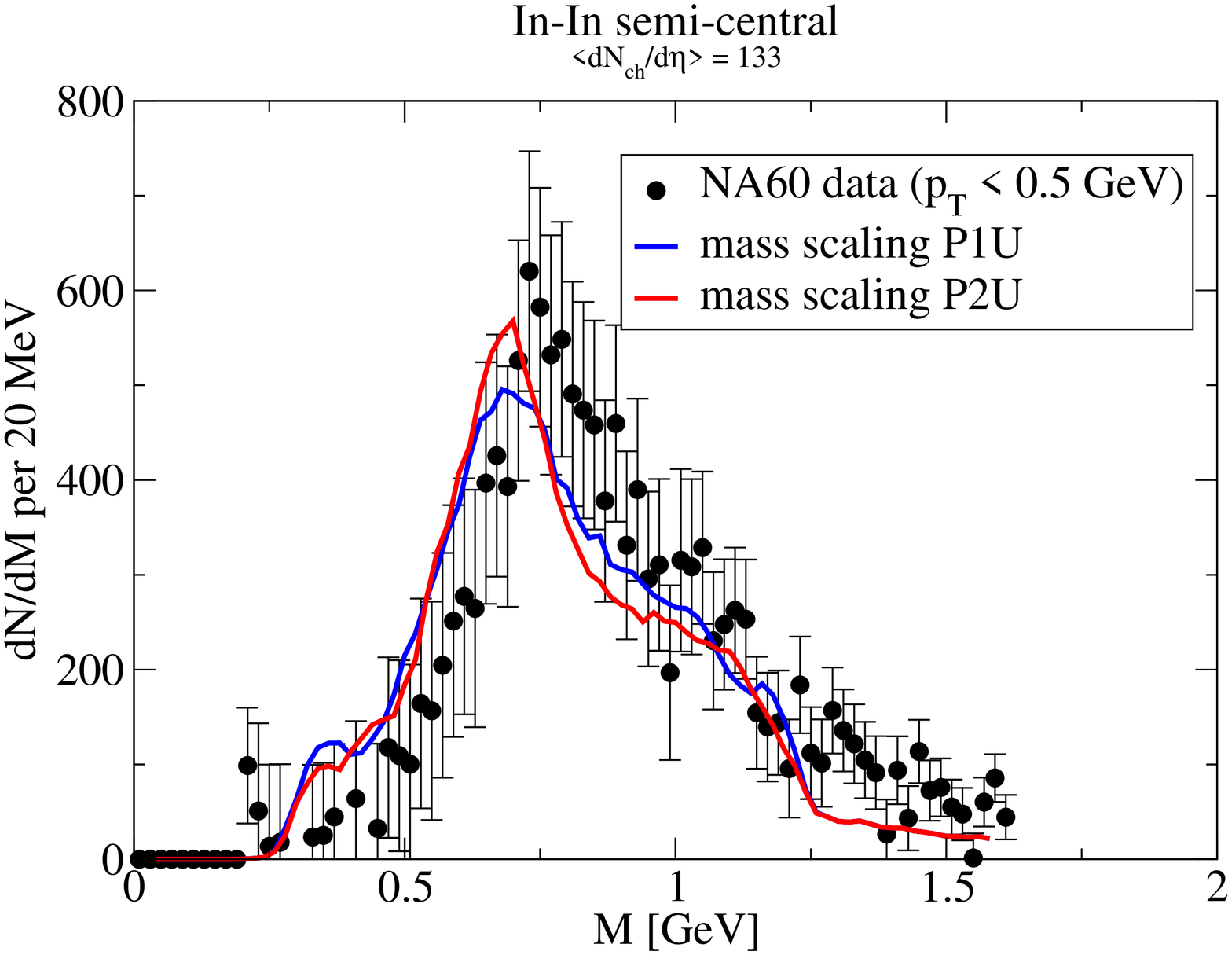, width=7cm,angle=-90}
\caption{\label{F-BR}Comparison of the NA60 dimuon data \cite{NA60data} for $p_T < 0.5~{\rm GeV}$  with schematic model calculations of Brown-Rho scaling in combination with the broadening compatible with QCD sum rules as described in \cite{BR}. Left panel: lower limits of the width for two parameter sets P1 to P2 describing the density dependence of the condensate. Right panel: upper limits of the width as determined by sum rules. The 3-momentum dependence of the spectral function has been neglected.
All model calculations have been normalized to the data in the mass interval $M < 0.9$ GeV.
}
\end{figure*}

We first present a comparison with the data using the spectral function for the $\pi-\rho$
systems for all $p_T$. The calculation contains a vacuum $\rho$ contribution
using a Cooper-Frye prescription (for details see section II E of \cite{Renk:2006dt}).
Most of the vacuum $\rho$-contribution is generated during the final breakup of the fireball \cite{Renk:2006ax,Renk:2006dt} and thus the underlying flow is strong and shuffles strength 
to higher $p_T$. 
This is the reason that the vacuum $\rho$-contribution is almost negligible for the $p_T<0.5~{\rm GeV}$ spectra, but is important for integrated $p_T$ (and high $p_T>1 {\rm GeV}$).  
The relative contribution of the vacuum $\rho$ and the in-medium $\rho$ appears to 
be approximately correct. While the strength of the vacuum $\rho$ parametrically scales like the freeze-out volume times the $\rho$-meson's life-time, the in-medium 
contribution is sensitive to the lifetime of the fireball. Thus the good description in the mass region $M \approx 0.6 - 0.9 {\rm GeV}$ is due to contributions of the vacuum and 
in-medium $\rho$ and therefore a consistency check of the fireball evolution.
We note that the $p_T>{1 GeV}$  spectra are 
 in good agreement with the NA60 data (comparison not shown).

The good agreement of the $\Phi$-derivable approach calculation with the data including the vacuum $\rho$  is to some degree surprising and unexpected, as the model contains the self-consistent broadening due to the $\pi-\rho$ interaction \cite{RR,Note} but is completely insensitive to the presence of other hadrons, especially baryons (which, based on perturbation theory, are expected to play a major role in the broadening of the $\rho$). 
One may take this as an indication that the generic shape of the spectral function and its 3-momentum dependence is right, not as a proof that this is the main mechanism capable of producing such a spectral function or that the dynamics of baryons is unimportant. However, these results clearly indicates that it is necessary to extend studies on non-perturbative mechanisms for broadening in the future.

The other model calculations (chiral model, Walecka model, mass scaling scenarios)
rely on the assumption of a $\rho$-meson at rest relative to the thermal heat bath 
${\bf q}=0$, hence they can only be compared to the data in the $p_T<0.5 {\rm GeV}$ cut, see \cite{Renk:2006dt}. This is somewhat unfortunate since the acceptance of the NA60's experimental setup is relatively small in the $M<0.6 
~{\rm MeV}$, $p_T<0.5 {\rm GeV}$ region, nonetheless interesting conclusions can be drawn.

The chiral model and the Walecka model in Fig. \ref{F-RR}, give a fair description of the data in the low $p_T$ region.
Although we note that the chiral model calculation also contains an in-medium $\phi$ and $\omega$ contribution (as these have different mass/width in medium than in vacuum such a contribution would not be removed by the experimental subtraction of the cocktail $\phi,\omega$) and apparently the strong in-medium $\phi$ broadening is not in good agreement with the low-$p_T$ data.

We present the results of the calculation with the mass scaling scenarios as described in \ref{mass-scale} in Fig. (\ref{F-BR})  after acceptance folding and maintaining the normalization for the two different parameter sets P1 \cite{Hatsuda:1991ez} and P2 \cite{Klingl:1997kf} used in the QCD sum rule approach \cite{Leupold:1997dg,BR}.
 Since the spectral function is  determined for vanishing three momentum, only comparison with the low $p_T$ binned data is meaningful. Given this comparison one realizes that such a $\rho$-mass scaling behavior is disfavored in the face of the experimental data. 
Although part of the observed enhancement below the mass of the free $\rho$-meson is accounted  for, the yield in the peak region is underpredicted.
As discussed above the vacuum $\rho$-contribution is not strong enough to account for this difference. 
What also can be stated is that scaling scenarios which would predict a dropping 
mass of the $\rho$-meson with the quark condensate that is much stronger 
than in the $a=1/2$ case, i.e.  $a \gg 1$, are in disagreement with the data. This has two reasons, namely, (i) they predict a stronger decrease of the $\rho$-meson mass with density leading to a stronger shift of the "peak" region down to small invariant masses of the di-muon spectra, and (ii) a stronger drop of the mass demands a much less rising or for $a>1$ even decreasing in-medium width of the $\rho$-meson with density \cite{BR}, both in disagreement with the NA60 dimuon data.
We note that Rapp and van Hees have studied a mass scaling model \cite{vanHees:2006iv} recently, where they use a mass parameterization of the type 
\begin{eqnarray}
m^*_{\rho}=m_\rho(1-0.15 \rho_B/\rho_0)[1-(T/T_c)^2]^{0.3}
\end{eqnarray} 
and employed vector meson dominance. They supplemented the in-medium spectral function by thermal broadening and obtained qualitatively similar results for mass-scaling scenarios, na\-mely somewhat an underprediction of the yield  in the peak region due to shift to lower masses.
We mention that Greiner and Schenke recently argued that non-equilibrium effects can lead to significant different results in comparison to equilibrium calculations due to the importance of memory effects \cite{Schenke:2005ry}. They find that this might have significance for the normalization and line-shape of dilepton spectra as obtained in mass scaling scenarios \cite{Greiner:2006ug}.

\vspace{-0.6cm}
\section{Conclusion}
\vspace{-0.2cm}

In these proceedings we have shown how a dynamical evolution model of $158~{\rm AGeV}$ In-In collisions can be constructed in order to calculate the measured di-muon yield by applying scaling arguments 
to an evolution model that is able to describe the  plethora of observables in Pb-Pb collisions \cite{SPS}.

Furthermore we discussed how the low-mass NA60 di\-muon measure\-ments can be used to disentangle different scenarios:  strong broadening of the spectral function with a small shift of the $\rho$-meson mass in medium as inferred by hadronic-many body calculations vs. considerable mass shift of the $\rho$-meson.
We have shown that taking into account only a mass shift without considering
changes of the in-medium width in a nuclear medium violates QCD sum rule constraints 
\cite{BR}. We therefore presented a schematic mass scaling model in order to test mass scaling assumptions in a dense medium taking the correlation between mass shifts and
finite width effects into account.
To account for the intrinsic uncertainties we used two different parameterization in the QCD sum rule approach. 
It was shown that the evolution model together with hadronic many body calculations which include considerable broadening can account for the data whereas the same evolution model assuming a  mass scaling scenario underpredicts the yield in the "peak" region of the data  and indicates a shift of the maximum of the di-muon spectrum down to smaller invariant masses. 

The NA60 data therefore do not favor downward mass scaling and clearly indicate that the dominant in-medium effect in the medium created in $158 {\rm AGeV}$ In-In collisions is broadening of the $\rho$-spectral function.

\section{Acknowledgment}

We like to thank Berndt M\"uller, Carlos Lourenco, Wolfram Weise and Norbert Kaiser for comments and discussions. We are grateful to Sanja Damjanovic for folding our model into  the  
NA60 acceptance simulation and thank her and Hans Specht  for comments and  discussions.
This work was supported by DOE grant DE-FG02-96ER40945 and the Alexander von Humboldt Foundation's Feodor Lynen Fellow program.

\vspace{-0.4cm}
%
\bibliographystyle{h-physrev3}
\bibliography{hotBIB}
%

\end{document}